\documentclass[cite]{epl}
\usepackage{amsmath}

\title{Full counting statistics of incoherent Andreev transport}
\shorttitle{FCS of incoherent Andreev transport}
\author{W. Belzig\inst{1} \and P. Samuelsson\inst{2}} 
\institute{
  \inst{1} Departement Physik und Astronomie, Universit\"at Basel,
  Klingelbergstrasse~82,
  CH-4056 Basel, Switzerland\\
  \inst{2} D\'epartement de Physique Th\'eorique, Universit\'e de
  Gen\`eve, CH-1211 Gen\`eve 4, Switzerland.} 


\begin{document}

\maketitle

\begin{abstract}
  We study the full counting statistics of heterostructures consisting
  of normal metal parts connected to a superconducting terminal.
  Assuming that coherent superconducting correlations are suppressed
  in the normal metals we show, using Keldysh-Nambu Green's function,
  that the system can be mapped onto a purely normal system with twice
  the number of elements. For a superconducting beam splitter with
  several normal terminals we obtain general results for the counting
  statistics.
\end{abstract}

\section{Introduction}

A complete statistical description of a transport process can be
obtained from the full counting statistics (FCS)
\cite{levitov:93-fcs}. The FCS of charge transport in mesoscopic
conductors has recently attracted a lot of attention
\cite{nazarov:03}. It was shown that scattering between uncorrelated
Fermi leads is described by binomial statistics
\cite{levitov:93-fcs,levitov:96-coherent}, in which charge is
transfered in elementary units. Subsequently, the FCS in
normal-superconducting systems have been studied and it was shown that
Andreev reflection leads to a doubled charge transfer
\cite{muzy:94,dejong:94,nagaev:01}.  Later on, several approaches to
FCS have been put forward. A theory of full counting statistics based
on Keldysh-Green's function was developed\cite{yuli:99-annals}. This
formulation allows a straightforward generalization to superconductors
\cite{belzig:01-super,belzig:01-diff}, multi-terminal structures
\cite{yuli:02-multi} and Coulomb blockade systems \cite{bagrets:02}.
As another example, the FCS of charge transport was expressed via the
counting statistics \cite{beenakker:02} of photons emitted from the
conductor. Very recently, a classical path-integral approach to FCS
was developed \cite{pilgram:03}.

The second moment of the FCS is related to the shot noise
\cite{blanter:00}. Suppression of the shot noise from the classical
Poisson value can result from the Fermi statistics of the electrons
\cite{khlus:87,lesovik:89,buettiker:92}. Particularly interesting are
the correlations between currents at different terminals -- the cross
correlations. For uncorrelated Fermi leads they are in general
negative \cite{buettiker:90,buettiker:92}, which has been
experimentally confirmed\cite{henny:99,oliver:99,oberholzer:00}. If
the current is injected from a superconducting terminal, the sign of
the cross-correlations may become positive
\cite{martin:96,anantram:96,torres:99,taddei:02,boerlin:02,samuelsson1}.
If the proximity effect is absent both negative cross
correlations\cite{gramesbacher:00,nagaev:01} and positive cross
correlations\cite{samuelsson2} were theoretically predicted.

The positive correlations are a direct consequences of the charge
being transfered in pairs across the normal-superconducting interface
\cite{muzy:94,boerlin:02,samuelsson1,samuel:03}. This is the case,
independent of the presence or absence of the proximity effect in the
normal metal. Since the proximity effect introduces additional pair
correlations, it is of special interest to investigate, under general
circumstances, the charge transfer in normal-superconducting systems
with suppressed proximity effect.

\section{Circuit Theory of Incoherent Andreev Transport}

All transport properties of quasiclassical super\-con\-duc\-tor-\-
normal metal heterostructures can be calculated from the so-called
circuit theory
\cite{yuli:94-circuit,yuli:99-supplat,yuli:99-annals,yuli:02-multi,belzig:03-book}.
This theory is a discretization of the normal metal parts into
connectors and nodes. The mesh has to be chosen in a way that
approximates the real structure to the required precision. To each
node (label i) a matrix Green's function $\check G_i$ is ascribed. A
matrix current $\check I_{ij}$ flows through the connector between
nodes i and j. Conservation of the matrix current on each node
$\sum_j\check I_{ij}=0$ together with the normalization condition
${\check G}_i^2=1$ determine the unique solution for the whole
circuit. Connections to the external world are represented by
terminals (labeled with $\alpha$) with a fixed matrix Green's function
$\check G_\alpha$.

Electrons and holes (connected by Andreev reflection at the
superconducting terminals) propagate through the structure.  In the
circuit theory the decoherence due to the finite energy difference
$2E$ between electron and hole is described by a \emph{decoherence
  terminal} $d$, which is connected to the respective node
\cite{yuli:99-supplat}. The matrix current from the node into the
decoherence terminal has the form ${\check I}_{id}=-i (e^2/\hbar)
(E/\delta) [\check \sigma_z,\check G_i]$, where $\delta$ is the
spacing.  Alternatively the coherence is suppressed by a magnetic flux
$\Phi$ penetrating the node. The corresponding matrix current reads $
\check I_i\sim e^2\left(\Phi/\Phi_0\right)^2 \left[ \check\sigma_z
  \check G_i\check\sigma_z,\check G_i\right]$.  Here $\check \sigma_z$
denotes the third Pauli matrix in the Nambu matrix space and $\Phi_0$
is the flux quantum.

The full matrix current conservation on node $i$ then takes the from
\begin{equation}
  \label{eq:matrix-current-conservation}
  \sum_j \check I_{ij}+\check I_{id}=0\,.
\end{equation}
If we assume that the relevant energies are large in comparison to the
mean level spacing $\delta$ multiplied by the dimensionless
conductances of the connectors to the node (or if the magnetic field
is large), all off-diagonal elements of $\check G_i$ in the Nambu
space are suppressed.  To show this, we choose a representation of the
Green's functions, in which the electron- and hole-Keldysh Green's
functions are block-diagonal. For example, a normal terminal (labeled
$\alpha$) has the form
\begin{equation}
  \label{eq:greensfunction2}
  \check G_{\alpha}=
  \left(
    \begin{array}[c]{cc}
      \hat G^e_{\alpha}(\chi_\alpha) & 0 \\
      0 & \hat G^h_{\alpha}(\chi_\alpha)
    \end{array}
  \right)\,.
\end{equation}
Here electron (hole) Green's function are still matrices in Keldysh
space, defined as
\begin{equation}
  \hat G^{e(h)}_{\alpha}(\chi_\alpha)=\pm e^{\pm
  i\chi_\alpha\hat\sigma_z} \left( \begin{array}[c]{cc}
  1-2f^{e(h)}_{\alpha} & -2f^{e(h)}_{\alpha} \\ -2(1-f^{e(h)}_{\alpha}) &
  2f^{e(h)}_{\alpha}-1 \end{array} \right) e^{\mp
  i\chi_\alpha\hat\sigma_z} \, 
\end{equation}
with $f^{e(h)}_{\alpha}(E)=(1+\mbox{exp}[(E\mp eV_{\alpha})/k_BT])^{-1}$. The
Green's function of the node $i$ is
\begin{equation}
  \label{eq:node}
  \check G_i=
  \left(\begin{array}[c]{cc}
    \hat G_i^e & \hat F_i^1\\
    \hat F_i^2 & \hat G_i^h
  \end{array}\right)\,.
\end{equation}
The decoherence matrix current is
\begin{equation}
  \label{eq:decoherence}
  \check I_{id} \sim
  \left[
    \left(\begin{array}[c]{cc}
      1 & 0 \\ 0 & -1
    \end{array}\right)\,,\,
    \left(\begin{array}[c]{cc}
      \hat G_i^e & \hat F_i^1\\
      \hat F_i^2 & \hat G_i^h
    \end{array}\right)
  \right]
  = 2
  \left(\begin{array}[c]{cc}
    0 & \hat F_i^1 \\
    \hat F_i^2 & 0
  \end{array}\right)\,.
\end{equation}
Thus, if the decoherence current dominates over all other currents,
the matrix current conservation equation
Eq. (\ref{eq:matrix-current-conservation}) becomes $\check
I_i^{dec}=0$ and the off-diagonal components of $\check G_i$ vanish,
\textit{i.~e.} $\check G_i$ becomes block-diagonal. 

The connection between two nodes is described in general by a matrix
current
\begin{equation}
  \label{eq:arbitrary-connector}
  \check I_{ij} = -\frac{2e^2}{h} \sum_n
  \frac{T_n\left[\check G_i,\check G_j\right]}{
    4+T_n\left(\left\{\check G_i,\check G_j\right\}-2\right)}\,,
\end{equation}
where $\{T_n\}$ is the ensemble of transmission eigenvalues.  The
matrix current between two nodes with block-diagonal Green's functions
is also block diagonal. Each block has again the form
(\ref{eq:arbitrary-connector}).  Thus, in the whole circuit (except at
the superconducting terminals) the electron and hole blocks decouple.

At energies well below the superconducting gap, a superconducting
terminal at zero chemical potential has the Green's function
\begin{equation}
  \label{eq:super-terminal}
  \check G_S = 
  \left(\begin{array}[c]{cc}
    0 & \hat 1 \\
    \hat 1 & 0
  \end{array}\right)\,.
\end{equation}
The matrix current to the superconducting terminal has the form
(\ref{eq:arbitrary-connector}), where the Green's function on the
normal side has a block-diagonal form. Then, the block-diagonal
components of the current can be rewritten, after some algebra,
as
\begin{equation}
  \label{eq:snmatrix-simple}
  \check I_{Si}=
  \left(\begin{array}[c]{cc}
    \hat I_{Si}^{eh} & 0\\
    0 & -\hat I_{Si}^{eh}
  \end{array}\right)\,, \hspace{0.5cm} 
\hat I_{Si}^{eh} =   -\frac{2e^2}{h} \sum_n 
   \frac{R_n^A\left[\hat G_i^{e},\hat G_i^{h}\right]}{
    4+R_n^A\left(\left\{\hat G_i^{e},\hat G_i^{h}\right\}-2\right)}\,.
\end{equation}
Here we introduced the Andreev reflection probabilities
$R_n^A=T_n^2/(2-T_n)^2$. Thus, the superconductor matrix current
constitutes a (2$\times$2)-matrix current between electron and hole
Green's functions of the node. Note, that the 'transmission
probabilities' are given by the Andreev reflection probabilities.  We
conclude, that the transmission properties between electron and hole
blocks (mirrored at the superconductor) have \textit{exactly} the form
of a normal contact with the usual transmission probabilities $T_n$
replaced by $R_n^A$.

\begin{figure}
\onefigure[width=0.7\textwidth]{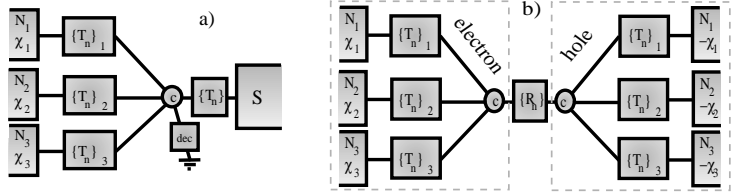}
\caption{Mapping of a beam splitter circuit. a) A number of normal
  terminal (here: three) is connected to one node. The node is
  connected to a superconducting terminal by a contact with
  transmission eigenvalues $\{T_n\}$. Strong decoherence of electrons
  and holes is accounted for by a decoherence terminal connected to
  the node. This allows to map the circuit on the structure depicted
  in b). All normal parts of the structure a) are doubled and the
  superconducting terminal connects these two circuits. The
  connector between electron and hole circuit has transmission
  eigenvalues $\{R_n^A\}$, where $R_n^A=T_n^2/(2-T_n)^2$ is the Andreev
  reflection probability corresponding to the transmission eigenvalues
  of the original circuit a).}
\label{fig:mapping}
\end{figure}

With this result, we are ready to present the mapping
rules\cite{sammap}, illustrated in Fig. \ref{fig:mapping}:
\begin{enumerate}
\item All normal elements, normal terminals and all connectors between
  normal parts of the circuit are doubled (electron and hole circuit). 
\item The superconducting terminals are the only connections between
  electron and hole circuit. They play the role of normal connectors,
  in which the normal transmission eigenvalues $T_n$ are replaced by
  the Andreev reflection probabilities $R_n^A=T_n^2/(2-T_n)^2$.
\item A counting field $\chi_\alpha^{e(h)}=\pm \chi$ is assigned to
  corresponding electron and hole terminals.
\item The counting statistics is obtained from the solution of the
normal circuit.
\end{enumerate}

An important general property of the extended circuit is electron-hole
symmetry. In our case this is formally a consequence of the perfect
symmetry of the extended circuit. Mathematically the symmetry relation
reads
\begin{equation}
  \label{eq:ehsym} 
  \hat\sigma_x\hat G_i^e (E) \hat \sigma_x = -\hat
  G_i^h (-E)\,.
\end{equation}
Below we will use this symmetry relation to obtain some general
properties of beam splitters.

We can draw a general conclusion about the parity of the transfered
charge number through the superconductor. Due to the Andreev process
this number should be even. Since we have replaced the superconducting
circuit by a purely normal on, the even parity is not obvious anymore.
However, for a two terminal device the FCS is $2\pi$-periodic in the
difference between the counting fields $\chi_e-\chi_h$, which,
according to our rules, should be replaced by $2\chi$. Thus, the CGF
is now a $\pi$-periodic function of $\chi$.  This argument can be
generalized to an arbitrary number of terminals, if we are interested
only in the charge transfer through the superconductor.  The total
transfered charge is obtained by taking all counting fields equal.
Again the CGF depends only on differences between counting fields and,
therefore, charge transfers between two electron (or hole) terminals
are not counted. With the same argument as given above, it follows
that the total CGF is invariant under a shift of all counting fields
by $\pi$.

\section{Counting statistics of two-terminal devices}
\label{sec:two-terminal}

As a check, we first apply the mapping rules to a simple two-terminal
contact without internal structure. The results for the cumulant
generating function (CGF) of a normal contact with transmission
eigenvalues $\{T_n\}$ is \cite{levitov:93-fcs} ($k_BT=0$)
\begin{equation}
  \label{eq:fcslevitov}
  S_{\textrm{NN}}(\chi_1,\chi_2) = 
  M\sum_n \ln\left[1+T_n(e^{i(\chi_1-\chi_2)}-1)\right]\,,
\end{equation}
where we have introduced the number of attempts per channel
$M=2eVt_0/h$.  Using the recipe outlined above we replace $T_n$ by
$R_n^A$ and $\chi_1-\chi_2$ by $2\chi$ and find
\begin{equation}
  \label{eq:fcsmuzkhm}
  S_{\textrm{SN}}(\chi)=M\sum_n 
  \ln\left[1+R_n^A(e^{i2\chi}-1)\right]\,,
\end{equation}
which is the result first obtained in Ref.~\cite{muzy:94}.

The counting statistics of a diffusive wire between a normal and a
superconducting terminal can also be found analytically. Let us write
the CGF of a normal diffusive conductor with conductance $g_N$ as
\cite{yuli:99-annals,levitov:96-diffusive,belzig:03-book}
\begin{equation}
  \label{eq:fcs-diffusive-normal} S_{\textrm{NN}}(\chi_1,\chi_2)=\frac{g_N V
    t_0}{4e} \textrm{acosh}\left(2e^{i\chi}-1\right)\,.
\end{equation}
Now, according to the mapping rules, we have to replace the diffusive
connector by two diffusive wires in series. We can neglect the
interface resistance and use that the CGF for a series of
two diffusive conductors is the same as for a single diffusive
conductor. The total conductance is halved and we thus find
\begin{equation}
  \label{eq:fcs-diffusive-super}
  S_{\textrm{SN}}(\chi)=
  \frac{g_N V t_0}{8e} \textrm{acosh}\left(2e^{i2\chi}-1\right)\,.
\end{equation}
This result proves that the FCS of the diffusive SN-wire in the
incoherent regime is the same as in the coherent
regime\cite{belzig:03-book}, which has so far only been demonstrated
for the first two moments \cite{dejong:94,nagaev:01,belzig:01-diff}.

As another example we consider a chaotic dot, which is much stronger
coupled to the superconductor than to the normal terminal. In this
case, the connector between the electron- and hole node does not
contribute to the FCS (the Green's functions in the electron and the
hole dots are the same). We can then consider the equivalent circuit
with a single dot, symmetrically coupled to the electron and
hole terminals. The Green's functions of electron and
hole terminal are (for $|E|<eV$ and
$k_BT=0$) 
\begin{equation}
  \label{eq:terminalgf}
  \hat G^{e(h)}=\mp \hat \sigma_z - 
  (\hat\sigma_x\pm i\sigma_y) e^{\pm i\chi_{e(h)}}\,.
\end{equation}
For a moment we take independent counting fields in the electron and
hole terminal, and replace $\chi_{e(h)}\to \pm\chi$
only in the end. The solution for the central node is
\begin{equation}
  \hat G_c =
  -\frac{1}{Z}\left[
    \hat\sigma_x \left(e^{i\chi_e}+e^{-i\chi_h}\right)
    +i\hat\sigma_y \left(e^{i\chi_e}-e^{-i\chi_h}\right)\right]\,,
\end{equation}
where $Z=\sqrt{\exp(i\chi_e-i\chi_h)}$. Using this result, we find the CGF
\begin{equation}
  S(\chi_e,\chi_h)=M\sum_n
  \ln\left[1+\frac{T_n}{2}
    \left(\sqrt{\exp(i\chi_e-i\chi_h)}-1\right)\right]\,.
\end{equation}
At this stage we can safely take the limit $\chi_{e(h)}=\pm\chi$ and
obtain
\begin{equation}
  S(\chi_e,\chi_h)=M\sum_n
  \ln\left[1+\frac{T_n}{2}
    \left(e^{i(\chi \textrm{mod} \pi)}-1\right)\right]\,,
\end{equation}
which ensures the $\pi$-periodicity of the CGF.

We now turn to the general case of the counting statistics of a beam
splitter as depicted in Fig.~\ref{fig:mapping}. The central node is
connected to several normal terminals (labeled with $\alpha$) and one
superconducting terminal. This structure has been studied in different
limits in Refs.~\cite{taddei:02,boerlin:02,samuelsson1,samuelsson2}.
From Eq.~(\ref{eq:snmatrix-simple}) it is clear that the Andreev
conductance $g_A=(4e^2/h) \sum_n R^A_n$ governs the coupling between
the electron and hole circuit.

\section{Counting statistics of a beam splitter -- weakly coupled
  superconductor} 
\label{sec:weak}

We assume here that the superconductor is only weakly coupled to the
beam splitter in the sense that $g_A\ll g_\Sigma\equiv \sum_\alpha
g_\alpha$, where $g_\alpha=(2e^2/h)\sum_n T_n$ is the conductance of
the connector to terminal $\alpha$. All normal terminals are held at
the same potential $V$ and zero temperature. Expanding all quantities
to first order in $g_A/g_\Sigma$, we find (similar to Ref.
\cite{samuelsson2}) that the total CGF can be expressed by the CGF of
the superconducting contact alone:
\begin{equation}
  \label{eq:weakS-fcs}
  S=M \sum_n \textrm{Tr}\ln \left[1-\frac{R^A_n}{4} 
    \left(\left\{\hat G^{e0}_c,\hat\sigma_x\hat G^{e0}_c\hat 
        \sigma_x\right\}+2\right)\right]\,.
\end{equation}
Here $\hat G^{e0}_c$ is the electron Green's function of the central
node in the absence of the superconductor (for $g_A=0$). All Green's
functions are evaluated at $E<V$ and we used Eq.  (\ref{eq:ehsym}) to
express S in terms of $\hat G^{e0}_c$ only. The Green's function $\hat
G^{e0}_c$ can be obtained quite generally. Due to the triangular shape
of all $\hat G^e_{\alpha}$ (see Eq.~(\ref{eq:terminalgf})), also $\hat
G^{e0}_c$ has the same form. The matrix current between the central
node and terminal $\alpha$ then becomes $\hat
I_\alpha=\frac{g_{\alpha}}{2} [\hat G^{e0}_c,\hat G^e_{\alpha}]$.
Thus, all normal connectors behave as tunnel contacts. The solution
for the central node is \cite{boerlin:02}
\begin{equation}
\label{eq:greenfcn}
 \hat G^{e0}_c=\frac{1}{g_\Sigma}\sum_\alpha g_\alpha\hat G^e_{\alpha} =
  \hat \sigma_z+(\hat\sigma_x+i\hat\sigma_y)\Lambda\,,
\end{equation}
where $\Lambda=\sum_\alpha p_\alpha e^{i\chi_\alpha}$
with $p_\alpha=g_\alpha/g_\Sigma$. It then
follows from Eq. (\ref{eq:weakS-fcs}), that the CGF is
\begin{equation}
  \label{eq:cgf_weakS}
  S(\{\chi_\alpha\})=M \sum_n\ln\left[1+R_n^A\left(\Lambda^2-1\right)\right]\,.
\end{equation}
Such a CGF leads to a FCS of the form
\begin{equation}
  P(N_1,N_2,...) = P_C(2Q) \frac12\left(1+\left(-1\right)^{2Q}\right)
  \frac{(2Q)!}{N_1!N_2!\cdots} p_1^{N_1} p_2^{N_2}\cdots
\end{equation}
where $Q=\sum_\alpha N_\alpha/2$ is the number of transfered Cooper
pairs.  The total probability distribution is therefore the
probability $P_C(2Q)$ that $Q$ Cooper pairs are transfered, and then
distributed among the normal terminals with respective probabilities
$p_\alpha$. A similar result has previously been obtained for tunnel
contacts \cite{boerlin:02} and a chaotic cavity \cite{samuelsson2}.
Here, this derivation holds for \textit{any} type of connector,
\textit{i.~e.}  point contacts, diffusive wires, single transparency,
etc.

We note that if the beam splitter is connected to the superconductor
by another connector (incoherent!), the solution is again given by
Eq. (\ref{eq:cgf_weakS}). This is the case since the additional node,
connected directly to the superconductor, has the same Green's
function as the central beamsplitter node, $\hat G^{e0}_c$, to leading
order in $g_A/g_\Sigma$, \textit{i.~e.} with the superconductor completely
decoupled.

\section{Counting statistics of a beam splitters -- strongly coupled
  superconductor} 
\label{sec:strong}

Another important case is when the superconductor is strongly coupled
to the beam splitter, \textit{i.~e.} $g_A \gg g_\Sigma$. In this limit the
superconducting connector is absent and the system is a chaotic dot
coupled to the electron and hole terminals.  We consider here the
case, in which the normal terminals are coupled by tunnel
contacts. The solution for the central node is then \cite{boerlin:02}
\begin{equation}
  \hat G_c=\hat K/\sqrt{{\hat K}^2}\quad,\quad
  \hat K=\sum_\alpha \frac{g_\alpha}{2} \left(\hat G^e_{\alpha}+\hat
  G^h_{\alpha}\right)\,.
\end{equation}
The CGF follows straightforwardly, 
\begin{equation}
  \label{eq:cgf-tunnel}
  S(\{\chi_\alpha\})=M \left(\textrm{Tr}\sqrt{{\hat
        K}^2}-g_\Sigma\right)
  =M \left[\sqrt{\left(\sum\nolimits_\alpha g_\alpha e^{i\chi_\alpha}\right)^2}-g_\Sigma\right]\,.
\end{equation}
In the present form the $\pi$-periodicity is evident. However,
calculating explicitly the cumulants by taking successive derivatives
of $S(\{\chi_\alpha\})$ with respect to $\chi_\alpha$, we can equally
well take the square-root in Eq.~(\ref{eq:cgf-tunnel}), giving
$S(\{\chi_\alpha\})=M\sum_\alpha
g_\alpha[\exp(i\chi_\alpha)-1]=\sum_\alpha S_\alpha(\chi_\alpha)$.
From this form of $S$, no longer evidently pi-periodic, it follows,
interestingly, that all cross correlators vanish.

\section{Conclusions}

We have studied the full counting statistics of normal
metal-superconductor heterostructures in the incoherent regime. The
original circuit with one superconducting terminal can be mapped on a
purely normal circuit consisting of an electron and a hole block. The
superconductor plays the role of a normal connector between the
electron and the hole block, with the usual transmission probabilities
$\{T_n\}$ replaced by the Andreev reflection probabilities
$\{R_n^A=T_n^2/(2-T_n)^2\}$. We have illustrated our approach with
several examples.


\acknowledgments

W. B. acknowledges support by the SNF and the NCCR
Nanoscience. P.S. acknowledges support by MaNEP.


\begin{thebibliography}{99}
\bibitem{sammap}
  Such a mapping was used for the second moment, the current
  correlations, in Ref.~\cite{samuelsson2}
\end{thebibliography}
\end{document}